\documentclass[twocolumn,prl]{revtex4}%
\usepackage{amsfonts}
\usepackage[T1]{fontenc}
\usepackage{amsmath,amsbsy,amssymb,graphicx}
\usepackage{amsmath}
\usepackage{amssymb}
\usepackage{graphicx}%

\begin{document}
\title{{\Large Dirac Fermions in Graphene Nanodisk and Graphene Corner: }\\Texture of Vortices with Unusual Winding Number}
\author{Motohiko Ezawa}
\affiliation{Department of Applied Physics, University of Tokyo, Hongo 7-3-1, 113-8656,
Japan }

\begin{abstract}
We analyze the zero-energy sector of the trigonal zigzag nanodisk and corner
based on the Dirac theory of graphene. The zero-energy states are shown to be
indexed by the edge momentum and grouped according to the irreducible
representation of the trigonal symmetry group $C_{3v}$. Wave functions are
explicitly constructed as holomorphic or antiholomorphic functions around the
K or K' point. Each zero-energy mode is a chiral edge mode. We find a texture
of magnetic vortices. It is intriguing that a vortex with the winding number
$2$ emerges in the state belonging to the $E$ (doublet) representation. The
realization of such a vortex is vary rare.
\end{abstract}
\maketitle

\begin{figure}[t]
\centerline{\includegraphics[width=0.44\textwidth]{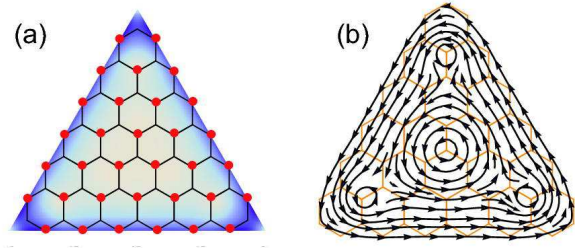}}\caption{(Color
on line) Trigonal zigzag graphene nanodisk. (a) The nanodisk size is defined
by $N=N_{\text{ben}}-1$ with $N_{\text{ben}}$\ the number of benzenes on one
side of the trigon. Here, $N=5$. The A sites are indicated by red dots. The
electron density is found to be localized along the edges. (b) A vortex
texture emerges in the real-space Berry connection. In this example, the
winding number is $2$ for the vortex at the center of mass while it is $1$ for
all others. }%
\label{FigContiDisk}%
\end{figure}

Graphene nanostructure\cite{GraphEx} has opened a field of carbon-based
nanoelectronics and spintronics alternative of silicon or GaAs. Carbon is a
common material and ecological. Large spin relaxation length is ideal for
spintronics\cite{Tombros}. The basic graphene derivatives are
nanoribbons\cite{Fujita96,EzawaRibbon,Brey73} and
nanodisks\cite{EzawaDisk,Fernandez,Hod,Wang,Potasz}. They correspond to
quantum wires and quantum dots, respectively. Regarding a nanodisk as a
quantum dot with internal degrees of freedom, similar but richer physics and
applications are anticipated\cite{EzawaCouloKondoSpin}. There are many type of
nanodisks, among which the trigonal zigzag nanodisk [Fig.\ref{FigContiDisk}]
is prominent in its electronic and magnetic properties owing to the
zero-energy sector\cite{EzawaDisk}. The main feature is that it becomes a
quasiferromagnet in the presence of Coulomb interactions\cite{EzawaDisk}.
Recently trigonal nanodisks have been experimentally obtained by way of the Ni
etching of a graphene sheet\cite{Ni}. An experimental realization will
undoubtedly accelerate further experimental and theoretical studies on
graphene nanodisks.

In graphene, the physics of electrons near the Fermi energy is described by
the massless two-component Dirac equation or the Weyl
equation\cite{Slonczewski,Semenoff,Ajiki}. Nanoribbons have been successfully
analyzed based on the Weyl equation\cite{Brey73}, but this is not yet the case
with respect to nanodisks. The Dirac theory of graphene nanodisks must be
indispensable to explore deeper physics and promote further researches.

The carbon atoms form a honeycomb lattice in graphene. We take the basis
vectors $\mathbf{a}_{1}=\left(  1,0\right)  a$\ and $\mathbf{a}_{2}%
=(1/2,\sqrt{3}/2)a$\ with $a$ the lattice constant ($a\approx2.46$\AA ). The
honeycomb lattice is bipartite and has two different atoms per primitive cell,
which we call the A and B sites [Fig.\ref{FigContiDisk}]. The Brillouin zone
is a hexagon in the reciprocal lattice with opposite sides identified. We
start with the tight-binding model (TBM) only with the nearest neighbor
hopping $t$. We ignore the spin degree of freedom in most parts in what follows.

The band structure is such that the Fermi point is reached by six corners of
the first Brillouin zone, among which there are only two inequivalent points.
We call them the K and K' points. The dispersion relation is linear around
them, $E_{\kappa}\left(  \mathbf{k}\right)  \simeq\hbar v_{\text{F}%
}|\mathbf{k}-\mathbf{K}_{\tau}|$ with $\tau=\pm$, where $v_{\text{F}}=\sqrt
{3}ta/2\hbar$ is the Fermi velocity and $\mathbf{K}_{\tau}=a^{-1}\left(
\tau2\pi/3,2\pi/\sqrt{3}\right)  $ for the K point ($K_{+}$) and the K' point
($K_{-}$).

The dispersion relation near the K and K' points is that of `relativistic'
Dirac fermions. Indeed, the TBM yields the quantum-mechanical
Hamiltonian\cite{Slonczewski,Semenoff,Ajiki},%
\begin{equation}
H_{\tau}=\hbar v_{\text{F}}\left(
\begin{array}
[c]{cc}%
0 & \tau\hat{k}_{x}-i\hat{k}_{y}\\
\tau\hat{k}_{x}+i\hat{k}_{y} & 0
\end{array}
\right)  ,\label{KP1}%
\end{equation}
where we have introduced the reduced wave number by $\hat{\mathbf{k}%
}=\mathbf{k}-\boldsymbol{K}_{\tau}$.\ The Hamiltonian acts on the
two-component envelope function, $\Phi_{\tau}=(\phi_{\text{A}}^{\tau}%
,\phi_{\text{B}}^{\tau})$. Each Hamiltonian describes the two-component
massless Dirac fermion or the Weyl fermion. The Weyl equations read%
\begin{equation}
i\hbar\partial_{t}\Phi_{\tau}(\mathbf{x})=v_{\text{F}}\mathbf{\sigma}%
\cdot\mathbf{p}_{\tau}\Phi_{\tau}(\mathbf{x}),\label{TimeWeyl}%
\end{equation}
where $\mathbf{p}_{\tau}=\hbar(\tau\hat{k}_{x},\hat{k}_{y})=-i\hbar
(\tau\partial_{x},\partial_{y})$. The wave function is given by $\psi
_{S}^{\tau}(\mathbf{x})=e^{i\boldsymbol{K}_{\tau}\cdot\boldsymbol{x}}\phi
_{S}^{\tau}(\mathbf{x})$.

The symmetries are as follows. We note that $H_{\text{K'}}=\sigma
_{y}H_{\text{K}}\sigma_{y}$ and $\sigma_{z}H_{\tau}\sigma_{z}=-H_{\tau}$,
where $\sigma_{y}$ and $\sigma_{z}$ are the generators of the mirror symmetry
and the electron-hole symmetry, respectively.

In terms of the complex variable, the Weyl equation reads
\begin{subequations}
\label{DiracEqGraph}%
\begin{align}
\partial_{z^{\ast}}\phi_{\text{A}}^{\text{K}}(\boldsymbol{x})  &
=i\varepsilon\phi_{\text{B}}^{\text{K}}(\boldsymbol{x}), & \partial_{z}%
\phi_{\text{B}}^{\text{K}}(\boldsymbol{x})  &  =i\varepsilon\phi_{\text{A}%
}^{\text{K}}(\boldsymbol{x}),\\
\partial_{z}\phi_{\text{A}}^{\text{K'}}(\boldsymbol{x})  &  =-i\varepsilon
\phi_{\text{B}}^{\text{K'}}(\boldsymbol{x}), & \partial_{z^{\ast}}%
\phi_{\text{B}}^{\text{K'}}(\boldsymbol{x})  &  =-i\varepsilon\phi_{\text{A}%
}^{\text{K'}}(\boldsymbol{x}),
\end{align}
with $\varepsilon=E/2\hbar v_{\text{F}}$. The envelope functions are
holomorphic or antiholomorphic for the zero-energy state ($E=0$).

\textit{Dirac electrons on zigzag edge:} We analyze a graphene sheet placed in
the upper half plane ($y\geq0$) with the edge at $y=0$. Translational
invariance in the $x$ direction dictates the envelope function is of the form
$\phi_{S}^{\tau}(x,y)=e^{i\hat{k}_{x}x}f_{S}^{\tau}\left(  y\right)  $. Due to
the analyticity requirement we obtain
\end{subequations}
\begin{subequations}
\begin{align}
\phi_{\text{A}}^{\text{K}}(\boldsymbol{x})  &  =C_{\text{A}}^{\text{K}%
}e^{i\hat{k}(x+iy)}, & \phi_{\text{B}}^{\text{K}}(\boldsymbol{x})  &
=C_{\text{B}}^{\text{K}}e^{i\hat{k}(x-iy)},\\
\phi_{\text{A}}^{\text{K'}}(\boldsymbol{x})  &  =C_{\text{A}}^{\text{K'}%
}e^{i\hat{k}(x-iy)}, & \phi_{\text{B}}^{\text{K'}}(\boldsymbol{x})  &
=C_{\text{B}}^{\text{K'}}e^{i\hat{k}(x+iy)},
\end{align}
with $C_{S}^{\tau}$ being normalization constants. Hereafter we ignore such
normalization constants.

According to the TBM result, there are no electrons in the B site on edges.
Hence we require $\phi_{\text{B}}^{\tau}\left(  y=0\right)  =0$. By avoiding
divergence at $y\rightarrow\infty$, the resultant envelope functions are found
to be $\phi_{\text{A}}^{\text{K}}(\boldsymbol{x})=e^{i\hat{k}z}$ for $\hat
{k}>0$ and $\phi_{\text{A}}^{\text{K'}}(\boldsymbol{x})=e^{i\hat{k}z^{\ast}}$
for $\hat{k}<0$, with all other components being zero.

The wave number is a continuous parameter for an infinitely long graphene
edge. According to the TBM result, the flat band emerges for
\end{subequations}
\begin{equation}
-\pi\leq ak<-2\pi/3\quad\text{and}\quad2\pi/3<ak\leq\pi,\label{RegionK}%
\end{equation}
around the K' and K points, respectively. The boundary points $ak=-\pi$ and
$ak=\pi$ are to be identified since they represent the same point in the
Brillouin zone.

\begin{figure}[t]
\centerline{\includegraphics[width=0.43\textwidth]{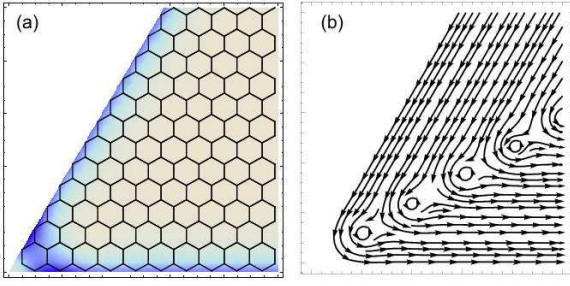}}\caption{(Color on
line) (a) Trigonal corner of graphene. The electron density is found to be
localized along the edges. (b) Real-space Berry connection for the trigonal
corner. A series of vortices are found to be present.}%
\label{FigCornerStream}%
\end{figure}

\textit{Dirac electrons in trigonal corner:} We apply the above analysis to
the study of the envelope functions for electrons in the zero-energy sector of
the zigzag trigonal corner [Fig.\ref{FigCornerStream}(a)], which is an
infinite region surrounded by the $x$ axis and the line with the angle $\arg
z=\pi/3$. They are holomorphic (antiholomorphic) around the K (K') point. Here
we discuss envelope functions around the K point. We start with the solution
$\phi_{1}\left(  z\right)  =Ce^{i\hat{k}z}$ for the upper half plane. We
rotate this by the angle $\pi/3$, which presents us with the solution
$\phi_{2}\left(  z\right)  =Ce^{i\hat{k}z\exp\left[  2\pi i/3\right]  }$ for
another half plane. The trigonal corner is given by the overlap region of
these two half planes, which is described by a linear combination of these two
functions with an appropriate coefficient. It is to be fixed by imposing the
boundary condition: Since the top of the corner is located at $z=0$, where
there is no atom, we impose $\phi_{\text{A}}^{\text{K}}(0)=0$. The resultant
function is $\phi_{\text{A}}^{\text{K}}(\mathbf{x})=\phi\left(  z\right)  $
with%
\begin{equation}
\phi\left(  z\right)  =e^{i\hat{k}z}-e^{i\hat{k}z\exp\left[  2\pi i/3\right]
}.
\end{equation}
The phase shift is $\pi$ at the corner. The envelope function around the K'
point ($\hat{k}<0$) is given by $\phi_{\text{A}}^{\text{K'}}(\mathbf{x}%
)=\phi\left(  z^{\ast}\right)  $.

We calculate the real-space Berry connection, $\mathcal{A}_{i}(\mathbf{x}%
)=-i\left\langle \phi\right\vert \partial_{i}\left\vert \phi\right\rangle $.
It exhibits a series of vortices [Fig.\ref{FigCornerStream}(b)], where the
wave function vanishes. It has\ an infinite number of zero points at
$z_{n}=(2\pi n/\sqrt{3}\hat{k})e^{\pi i/6}$, $n=1,2,3,\cdots$, around which it
is expanded as $\phi\left(  z\right)  =\hat{k}(z-z_{n})$.

\textit{Dirac electrons in trigonal nanodisk:} Our main purpose is to apply
the above result to the analysis of the zero-energy sector of the trigonal
zigzag nanodisk [Fig.\ref{FigContiDisk}]. The envelope function of the
trigonal zigzag nanodisk can be constructed by making a linear combination of
envelope functions for three trigonal corners. We consider the trigonal region
whose corners are located at $z_{1}=\left(  L,0\right)  $, $z_{2}=\left(
-L,0\right)  $, $z_{3}=\left(  0,\sqrt{3}L\right)  $. As the boundary
conditions we impose $\phi\left(  z_{1}\right)  =\phi\left(  z_{2}\right)
=\phi\left(  z_{3}\right)  =0$. The envelope function is obtained around the K
point ($\hat{k}>0$) as $\phi_{\text{A}}^{\text{K}}(\mathbf{x})=\phi\left(
z\right)  $ with
\begin{align}
\phi\left(  z\right)   &  =e^{i\hat{k}z}-e^{i\hat{k}L}e^{i\hat{k}\left(
z-L\right)  \exp\left[  -2\pi i/3\right]  }\nonumber\\
&  -e^{-i\hat{k}L}e^{i\hat{k}\left(  z+L\right)  \exp\left[  2\pi i/3\right]
}.\label{WaveFuncTrigo}%
\end{align}
The envelope function around the K' point ($\hat{k}<0$) is given by
$\phi_{\text{A}}^{\text{K'}}(\mathbf{x})=\phi\left(  z^{\ast}\right)  $. Note
that $\phi_{\text{B}}^{\tau}(\mathbf{x})=0$ identically, as is consistent with
the TBM result\cite{EzawaDisk}.

The wave number is quantized for a finite edge such as in the trigonal
nanodisk. We focus on the wave function $\psi_{\text{A}}^{\tau}\left(
\mathbf{x}\right)  $ at one of the A sites on an edge. For definiteness let us
take it on the $x$-axis. We investigate the phase shift between two points
$(x,0)$ and $(x+ma,0)$,
\begin{equation}
\Theta^{\tau}(x,m)=\tau\frac{2\pi m}{3}+\arg\phi\left(  x+ma\right)  -\arg
\phi\left(  x\right)  ,
\end{equation}
with (\ref{WaveFuncTrigo}). There are $N$ links along one edge of the size-$N
$ nanodisk, for which we obtain precisely $\Theta^{\tau}(a/2,N)=Nak$. On the
other hand, the phase shift is $\pi$ at the corner. The total phase shift is
$3Nak+3\pi$, when we encircle the nanodisk once. This phase shift agrees with
the TBM result. By requiring the single-valueness of the wave function, and
taking into account the allowed region of the wave number (\ref{RegionK}), we
find that the wave number is quantized as%
\begin{equation}
ak_{n}=\pm\left[  (2n+1)/3N+2/3\right]  \pi,\quad0\leq n\leq
(N-1)/2.\label{ValueK}%
\end{equation}
When $N$ is even, there are $N/2$ states for $k_{n}>0$ and $N/2$ states
$k_{n}<0$. When $N$ is odd, there are $(N-1)/2$ states for $k_{n}>0$ and
$(N-1)/2$ states for $k_{n}<0$. Additionally, there seem to appear two modes
with $ak_{n}=\pm\pi$ at $n=(N-1)/2$. However, they are identified with one
another, since they are located at the boundary of the Brillouin zone. There
are $N$ states in both of the cases, as agrees with the TBM
result\cite{EzawaDisk}.

\begin{figure}[t]
\centerline{\includegraphics[width=0.44\textwidth]{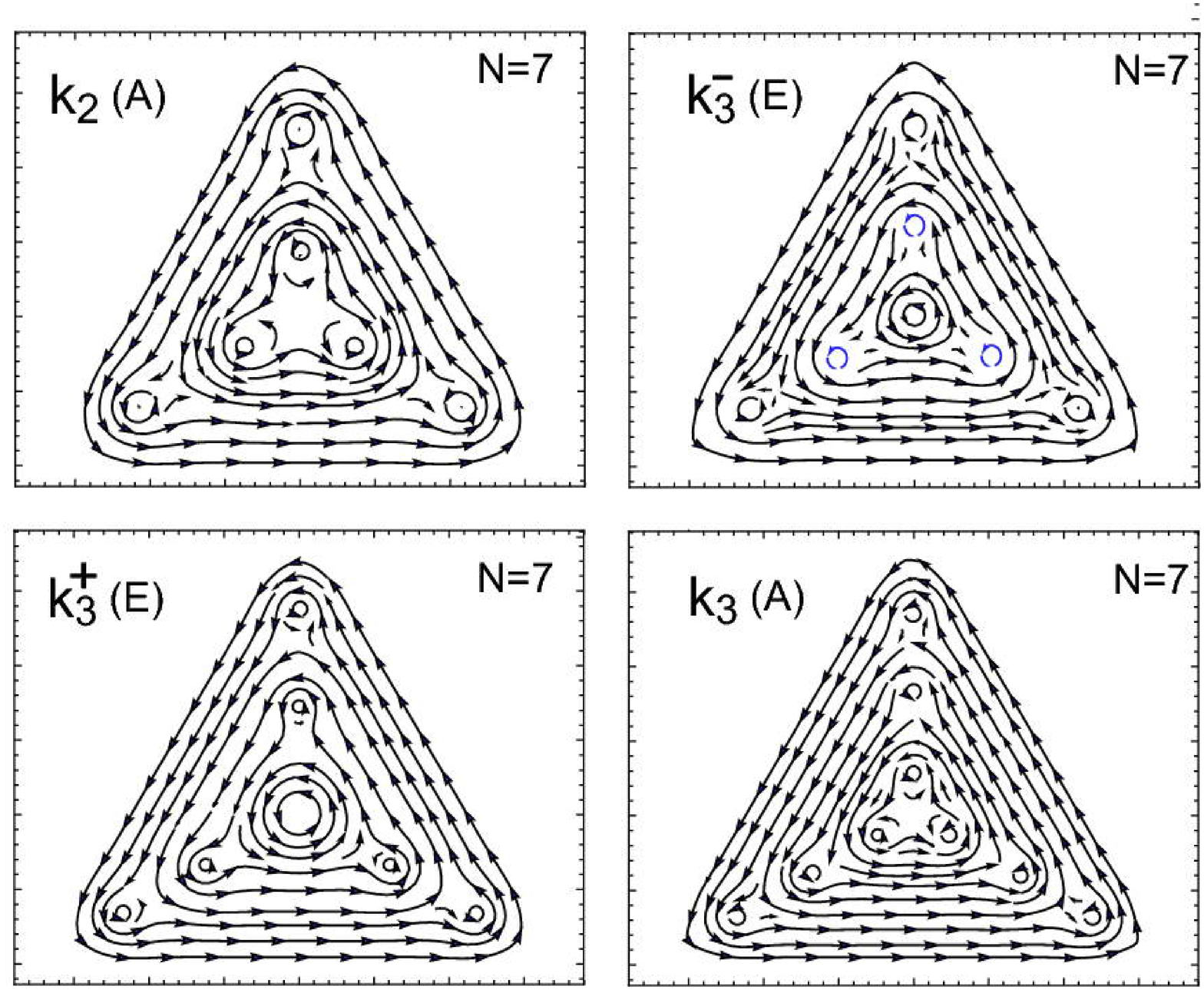}}\caption{Vortex
textures in the real-space Berry connection for the state $|k_{n}^{\alpha
}\rangle$ in the nanodisk with $N=7$. The representation is indicated in the
parenthesis. There are $n$ vortices along the $y$-axis in $|k_{n}^{\alpha
}\rangle$. A vortex appears at the center of mass for the $E$-mode
$|k_{n}^{\pm}\rangle$. It is interesting that the winding number is $2$ in the
state $|k_{n}^{+}\rangle$. }%
\label{FigBerryN7}%
\end{figure}

\textit{Trigonal symmetry group:} The symmetry group of the trigonal nanodisk
is $C_{3v}$, which is generated by the $2\pi/3$ rotation $\mathfrak{c}_{3}$
and the mirror reflection $\sigma_{v}$. It has the irreducible representation
\{$A_{1}$, $A_{2}$, $E$\}. The $A_{1}$ representation is invariant under the
rotation $\mathfrak{c}_{3}$ and the mirror reflection $\sigma_{v}$. The
$A_{2}$ representation is invariant under $\mathfrak{c}_{3}$ and antisymmetric
under $\sigma_{v}$. The $E$ representation acquires $\pm2\pi/3$ phase shift
under the $2\pi/3$ rotation. The $A_{1}$ and $A_{2}$ are 1-dimensional
representations (singlets) and the $E$ is a 2-dimensitional representation
(doublet). These properties are summarized in the following character table.%
\begin{equation}%
\begin{array}
[c]{c|ccc}\hline
C_{3v} & \mathfrak{e} & 2\mathfrak{c}_{3} & 3\sigma_{v}\\\hline
A_{1} & 1 & 1 & 1\\
A_{2} & 1 & 1 & -1\\
E & 2 & -1 & 0\\\hline
\end{array}
\end{equation}

The mirror symmetry is equivalent to the exchange of the K and K' points. With
respect to the rotation there are three elements $\mathfrak{c}_{3}^{0}$,
$\mathfrak{c}_{3}$, $\mathfrak{c}_{3}^{2}$, which correspond to $1$, $e^{2\pi
i/3}$, $e^{4\pi i/3}$. Accordingly, the phase shift of one edge is $0 $,
$2\pi/3$, $4\pi/3$. From this requirement we deduce that the state, indexed by
the edge momentum $k_{n}$ as in (\ref{ValueK}), is grouped according to the
representation of the symmetry group $C_{3v}$ as follows,
\begin{equation}%
\begin{array}
[c]{cc}%
\left.
\begin{array}
[c]{cc}%
A_{1}\text{ (singlet)}: & |k_{n}^{0}\rangle+|-k_{n}^{0}\rangle,\\
A_{2}\text{ (singlet)}: & |k_{n}^{0}\rangle-|-k_{n}^{0}\rangle,
\end{array}
\right\}  & \displaystyle k_{n}^{0}=\frac{6n+3}{3Na}\pi,\\%
\begin{array}
[c]{cc}%
E\text{ (doublet)}: & |k_{n}^{\pm}\rangle,\quad|-k_{n}^{\pm}\rangle,
\end{array}
& \displaystyle k_{n}^{\pm}=\frac{6n\pm1}{3Na}\pi,
\end{array}
\label{TrigoWave}%
\end{equation}
where $k_{n}^{\alpha}$ is subject to the condition (\ref{RegionK}). It follows
that
\begin{equation}
\left\lfloor (N+1)/3\right\rfloor \leq n\leq\left\lfloor N/2\right\rfloor ,
\end{equation}
where $\left\lfloor a\right\rfloor $ denotes the maximum integer equal to or
smaller than $a$. Some examples read%
\begin{equation}%
\begin{array}
[c]{c|c|c|c|c|c}%
N & 3 & 4 & 5 & 6 & 7\\\hline
A_{1} &  & k_{1}^{0} &  & k_{2}^{0} & k_{2}^{0}\\\hline
A_{2} & k_{1}^{0} & k_{1}^{0} & k_{2}^{0} & k_{2}^{0} & k_{2}^{0},k_{3}%
^{0}\\\hline
E & \pm k_{1}^{+} & \pm k_{2}^{-} & \pm k_{2}^{-},\pm k_{2}^{+} & \pm
k_{2}^{+},\pm k_{3}^{-} & \pm k_{3}^{-},\pm k_{3}^{+}%
\end{array}
\label{TrigonRepre}%
\end{equation}
The numbers of doublets ($E$-mode) and singlets ($A_{i}$-mode) are given by
$\left\lfloor \frac{1}{3}(N+1)\right\rfloor $ and $N-2\left\lfloor \frac{1}%
{3}(N+1)\right\rfloor $, respectively.

\textit{Berry connection:} To see the meaning of the wave number
$k_{n}^{\alpha}$ more in detail, we have calculated the Berry connection for
various states, which we show for the case of $N=7$ in Fig.\ref{FigBerryN7}.
Each mode is found to be chiral. We observe clearly a texture of vortices: The
number of vortices is $6,7,7,9$ for $|k_{2}^{0}\rangle$, $|k_{3}^{-}\rangle$,
$|k_{3}^{+}\rangle$, $|k_{3}^{0}\rangle$, respectively. The vortex at the
center of mass has the winding number $2$ in $|k_{3}^{+}\rangle$. In general,
the total winding number $N_{\text{vortex}} $ is calculated by%
\begin{equation}
N_{\text{vortex}}=\frac{-i}{2\pi}%
{\displaystyle\oint}
dx_{i}\,\frac{\phi^{\ast}(x,y)\partial_{i}\phi(x,y)}{|\phi(x,y)|^{2}%
}=N+m-1,\label{PhaseAB}%
\end{equation}
with $m=0,1,2,\cdots,\left\lfloor (N-1)/2\right\rfloor $ in the size-$N$
nanodisk, where the integration is made along the closed edge of a nanodisk.
There are $n$ vortices along the $y$-axis in the state $|k_{n}^{\alpha}%
\rangle$. The state $|k_{n}^{\pm}\rangle$, being the $E$-mode, has a vortex at
the center of mass, where the winding number is $2$ in the state $|k_{n}%
^{+}\rangle$. On the other hand, the state $|k_{n}^{0}\rangle$ does not have a
vortex at the center of mass, and the combinations $|k_{n}^{0}\rangle
\pm|-k_{n}^{0}\rangle$ belong to the A$_{1}$ and A$_{2}$ representations,
respectively. This statement is demonstrated by investigating the zero points
of the envelop function (\ref{WaveFuncTrigo}), where vortices appear. For
instance, it is expanded around the center of mass $z=z_{0}$ as $\psi\left(
z\right)  =\sum_{n=0}C_{n}(z-z_{0})^{n}$, where the coefficients $C_{0}$ and
$C_{1}$ are found to vanish at $k=k_{n}^{\pm}$ and $k=k_{n}^{+}$,
respectively, with $C_{3}\neq0$. Hence the winding number is $2$ for
$k=k_{n}^{+}$.

It should be emphasized that there exists a good agreement with respect to the
edge states between the Dirac description and the exact diagonalization
results of the TBM even for a small system. Indeed, we can easily compute the
phase at each lattice point by exact numerical methods. Then, comparing it
with the result due to the Dirac description, it is easy to see that a good
agreement holds between them. This shows that the real space Barry connection
computed by exact numerical methods has the same physical reality with the one
obtained by using the Dirac Hamiltonian.

\begin{figure}[t]
\centerline{\includegraphics[width=0.44\textwidth]{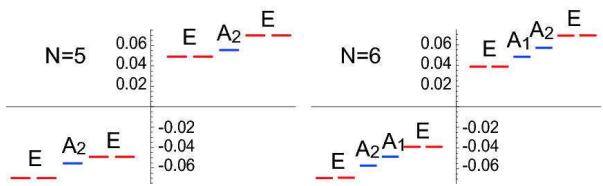}}\caption{(Color on
line) The energy spectrum with Coulomb interactions in nanodisks with $N=5,6$,
derived based on the tight-bindind model. The vertical axis stands for the
energy in unit of $1$eV. The $N$-folded degenerate states in the
noninteraction regime split into different levels according to the
representation of the trigonal symmetry $C_{3v}$, as indicated. For instance,
the positive (negative) energy levels are for down-spin (up-spin) states. The
ground state is a ferromagnet.}%
\label{FigSplit}%
\end{figure}

\textit{Zero-energy splitting due to Coulomb interactions:} We have
constructed explicitly the wave function for each state in the size-$N$
trigonal nanodisk. All these states are edge modes belonging to the
zero-energy sector. When Coulomb interactions are taken into account, the
degeneracy among the zero-energy states is resolved\cite{EzawaDisk}. The
Coulomb Hamiltonian has the trigonal symmetry $C_{3v}$, and the energy
spectrum splits into different levels according to its representation, as
illustrated in Fig.\ref{FigSplit}.

There exists additionally the spin degeneracy in the noninteracting
Hamiltonian: The total degeneracy is $2N$. The spin degeneracy is broken
spontaneously due to the exchange interaction when Coulomb interactions are
introduced\cite{EzawaDisk}. The splitting is symmetric with respect to the
zero-energy level. At half-filling, electrons with the identical spin fill all
energy levels under the Fermi energy. Then, the spin of the ground state is
$N/2$, and it is a ferromagnet. We show the energy spectrum for $N=5,6$ in
Fig.\ref{FigSplit}. By tuning the chemical potential any of them is made the
ground state.

\textit{Magnetic vortices:} The zero-energy degeneracy is resolved by the
Coulomb interaction, and the dispersion relation becomes nontrivial. The
time-dependent solution is well known,
\begin{subequations}
\begin{align}
\psi_{\text{A}}^{\text{K}}(t,\boldsymbol{x})= &  e^{-i(E_{n}^{\alpha}%
t-k_{n}^{\alpha}x)-k_{n}^{\alpha}y},\\
\psi_{\text{A}}^{\text{K'}}(t,\boldsymbol{x})= &  e^{-i(E_{n}^{\alpha}%
t+k_{n}^{\alpha}x)-k_{n}^{\alpha}y},
\end{align}
where $E_{n}^{\alpha}$ is the energy of the state $|k_{n}^{\alpha}\rangle$
[Fig.\ref{FigSplit}]. Here we have suppressed the spinor part. On one hand,
the $A_{i}$-modes $|k_{n}^{0}\rangle\pm|-k_{n}^{0}\rangle$ represent standing
waves. On the other hand, the $E$-modes $|k_{n}^{\beta}\rangle$ and
$|-k_{n}^{\beta}\rangle$ represent the right-propagating mode and the
left-propagating mode, respectively, for $\beta=\pm$.

Charged particles propagating along a closed edge generates magnetic field.
The electromagnetic interaction is described in terms of the electromagnetic
potential $\mathbf{A}$, which is introduced to the system by way of the
Peierls substitution $\partial_{j}\rightarrow\partial_{j}+ieA_{j}/\hbar$. From
the Weyl equation (\ref{DiracEqGraph}) we derive
\end{subequations}
\begin{equation}
eA_{i}(\boldsymbol{x})=\hbar\mathcal{A}_{i}(\boldsymbol{x})/|\phi_{\text{A}%
}^{\text{K}}(\boldsymbol{x})|^{2}%
\end{equation}
with $\mathcal{A}_{i}(x,y)=-i\phi_{\text{A}}^{\text{K}\ast}(\mathbf{x}%
)\partial_{i}\phi_{\text{A}}^{\text{K}}(\mathbf{x})$ in the lowest order of
approximation, where $\phi_{\text{A}}^{\text{K}}(\mathbf{x})$ is assumed to be
not modified from (\ref{WaveFuncTrigo}). The potential $A_{i}(\boldsymbol{x})$
exhibits the same texture of vortices as in Fig.\ref{FigBerryN7}. The magnetic
field is given by%
\begin{equation}
B(\boldsymbol{x})=\nabla\times\mathbf{A}(\boldsymbol{x})=\frac{2\pi\hbar}%
{e}\sum_{n}\nu_{n}\delta\left(  z-z_{n}\right)  ,
\end{equation}
where $\nu_{n}$ stands for winding number of the vortex at $z=z_{n}$. Hence a
texture of vortices in the Berry connection leads to a texture of magnetic
vortices. A comment is in order. This $\delta$-function type magnetic field
would be smoothed out in a rigorous analysis of the coupled system of the
Maxwell equation and the Weyl equation.

It is intriguing that, by tuning the chemical potential, a vortex with the
winding number $2$ emerges in the ground state $|k_{n}^{+}\rangle$. As is well
known, a single flux quantum has experimentally been observed in
superconductor by using an electron-holographic interferometry\cite{Tonomura}.
Then, in principle it is possible to observe a vortex texture in nanodisk as
well. Furthermore, by attaching a superconductor film one may observe a
disintegration of a vortex into two when the flux enters into the
superconductor from the nanodisk. This would verify the winding number 2 of a vortex.

\textit{Conclusions:} In this paper we have classified the zero-energy sector
of the trigonal zigzag nanodisk into a fine structure according to the
trigonal symmetry group $C_{3v}$. We have explicitly constructed wave
functions based on the Dirac theory and specified them by the quantized edge
momentum. A texture of magnetic vortices is found to emerge, which has an
unusual winding number. As far as we are aware of, the vortex with the winding
number 2 has never been found in all branches of physics. This is because two
vortices with the winding number 1 have lower energy than one vortex with the
winding number 2 in general. In the present case the disintegration of a
vortex into two is prohibited by the trigonal symmetry.

I am very much grateful to Professors N. Nagaosa and H. Tsunetsugu for
fruitful discussions on the subject and reading through the manuscript. This
work was supported in part by Grants-in-Aid for Scientific Research from the
Ministry of Education, Science, Sports and Culture No. 20940011.


\begin{thebibliography}{99}                                                                                               %
\bibitem {GraphEx}K. S. Novoselov, \textit{et al.}, Science \textbf{306}, 666
(2004). K. S. Novoselov, \textit{et al.}, Nature \textbf{438}, 197 (2005). Y.
Zhang, \textit{et al.}, Nature \textbf{438}, 201 (2005).

\bibitem {Tombros}N. Tombros, \textit{et al.}, Nature \textbf{448}, 571 (2007).

\bibitem {Fujita96}M. Fujita, \textit{et al.}, J. Phys. Soc. Jpn. \textbf{65},
1920 (1996).

\bibitem {EzawaRibbon}M. Ezawa, Phys. Rev. B, \textbf{73}, 045432 (2006).

\bibitem {Brey73}L. Brey, and H. A. Fertig, Phys. Rev. B, \textbf{73}, 235411 (2006).

\bibitem {EzawaDisk}M. Ezawa, Phys. Rev. B \textbf{76}, 245415 (2007): M.
Ezawa, Physica E \textbf{40}, 1421-1423 (2008): M. Ezawa, New J. Phys. 11,
095005 (2009).

\bibitem {Fernandez}J. Fern\'{a}ndez-Rossier, and J. J. Palacios, Phys. Rev.
Lett. \textbf{99}, 177204 (2007).

\bibitem {Hod}O. Hod, V. Barone, and G. E. Scuseria, Phys. Rev. B \textbf{77},
035411 (2008).

\bibitem {Wang}W. L. Wang, S. Meng and E. Kaxiras, Nano Letters \textbf{8},
241 (2008).

\bibitem {Potasz}P. Potasz, A. D. G\"{u}\c{c}l\"{u} and P. Hawrylak, Phys.
Rev. B \textbf{81}, 033403 (2010).

\bibitem {Ni}L.C. Campos et al., Nano Lett., \textbf{9}, 2600 (2009).

\bibitem {EzawaCouloKondoSpin}M. Ezawa, Phys. Rev. B \textbf{77}, 155411
(2008); \textit{ibid.} B \textbf{79}, 241407(R) (2009): M. Ezawa, Eur. Phys.
J. B \textbf{67}, 543 (2009)

\bibitem {Slonczewski}J.C. Slonczewski and P.R. Weiss, Phys. Rev.
\textbf{109}, 272 (1958).

\bibitem {Semenoff}G. W. Semenoff, Phys. Rev. Lett. \textbf{53}, 2449 (1984).

\bibitem {Ajiki}H. Ajiki and T. Ando, J. Phys. Soc. Jpn., \textbf{62}, 1255
(1993); T. Ando, Y. Zheng and H. Suzuura, Microelectron. Eng., \textbf{63},
167 (2002).

\bibitem {Tonomura}T. Matsuda, \textit{et al}., Phys. Rev. Lett. \textbf{62},
2519 (1989).
\end{thebibliography}
\end{document}